\documentclass[page-classic,figures]{epl2}   

\title{ Optical readout of charge and spin in a self-assembled   
  quantum dot in a strong magnetic field}   
\shorttitle{ Optical readout of charge and spin in a self-assembled 
  quantum dot in a strong magnetic field}

\author{M. Korkusinski\inst{1}
\and P. Hawrylak\inst{1}
\and A. Babinski\inst{2}
\and M. Potemski\inst{3}
\and S. Raymond\inst{1}
\and Z. Wasilewski\inst{1}}
\shortauthor{M. Korkusinski \etal}
   
\institute{
  \inst{1} Institute for Microstructural Sciences, National Research
  Council, Ottawa, Canada,  K1A0R6\\
  \inst{2} Institute of Experimental Physics, Warsaw University,
  Warsaw, Poland \\
  \inst{3} Grenoble High Magnetic Field Laboratory, CNRS Grenoble,   
  France
}   


\abstract{
We present a theory and experiment demonstrating optical readout of
charge and spin in a single InAs/GaAs self-assembled quantum dot. 
By applying a magnetic field we create the filling factor 2 quantum
Hall singlet phase of the charged exciton. 
Increasing or decreasing the magnetic field leads to electronic
spin-flip transitions and increasing spin polarization.
The increasing total spin of electrons appears as a manifold of
closely spaced emission lines, while spin flips appear as
discontinuities of emission lines.  
The number of multiplets and discontinuities measures the number of
carriers and their spin. 
We present a complete analysis of the emission spectrum of a single
quantum dot with $N=4$ electrons and a single hole, calculated and
measured in magnetic fields up to 23 Tesla.   
}

\begin{document}

\maketitle   
  
 
Electronic properties of quantum dots and their potential applications 
in nano-electronics, nano-spintronics, nano-optics and quantum  
information processing  depend critically on the number of   
confined carriers \cite{spintronics}.   
This number can be tuned with metallic gates or by optical  
means \cite{andy-filling,sad-filling}. 
In lateral gated quantum dots, the electron number can be determined
by attaching leads and counting the number of  spin flips in the
Coulomb blockade spectra \cite{gated-spinflips}.    
In a self-assembled quantum dot (SAD) \cite{michler-book}
electrons can be counted optically by adding a single
hole to the confined electrons and observing the emission spectrum of
the resulting charged exciton. 
This can be achieved by tuning the number of electrons in the SAD with
an electrostatic potential created by a gate, and comparing the
emission spectra obtained for different charge states of the system
\cite{chargedx-experim-gated}.
Here we demonstrate that the determination of charge is also possible
for a fixed number of electrons, without a need for attaching gates.
We achieve this by detecting the optical signatures of electronic spin
flips in an external magnetic field.
The spin-flip transitions require very high magnetic fields to resolve
the Landau-level structure of single-particle states \cite{hifield}. 
The optical measurement of the electron number is nontrivial as the
optical transitions involve only electrons from the lowest occupied
state and are, at first sight, an insensitive probe of the number of
carriers $N$ \cite{chargedx-experim-gated,chargedx-experim-mesas,manfred-fine,chargedx-theory}.
The proposed readout scheme is demonstrated on the example of the
emission spectra of the triply charged exciton $X3-$, composed of four
electrons and one valence hole, in magnetic fields of up to 23 T.    
We show that the magnetic field builds up the total electron spin
through spin flips.  
Different total spin states are characterized by different emission
spectra, including spin multiplets due to exchange interaction with
valence hole.  
This is analogous to an exciton interacting with the spin of a 
magnetic ion localized in a quantum dot \cite{besombes}, except that 
here the total spin of electrons is engineered by the magnetic 
field.   
 
  
The single-particle energies of an electron and a hole confined in a 
SAD and in a magnetic field $B$ parallel to the SAD's growth direction
are well approximated by 
those of a pair of harmonic oscillators \cite{hifield}     
$\varepsilon^{\beta}(n,m,\sigma) = \Omega_+^{\beta}\left(n + {1\over2}   
\right) + \Omega_-^{\beta}\left(m+{1\over2}\right)+g_{\beta}\mu_B
B\sigma_{\beta}$,
where $n,m=0,1,\dots$,  $\beta=e/h$ denotes the 
particle type, $g_{\beta}$ is the effective Land\'e factor, 
$\mu_B$ is the Bohr magneton, and $\sigma_{\beta}$ is the $z$ 
component of the spin ($\sigma_e=\pm1/2$,  
$\sigma_h=\pm3/2$).    
The oscillator energies $\Omega_{\pm}^{\beta}=\Omega_h^{\beta}\pm   
{1\over2}\Omega_{c,\beta}$, where $\Omega_{c,\beta}=eB/m^*_{\beta}$ is the   
cyclotron energy, $m^*_{\beta}$    is the effective mass,  
$\Omega_h^{\beta}=\sqrt{\Omega_{0,\beta}^{2}+{1\over4}\Omega_{c,\beta}^2}$   
and $\Omega_{0,\beta}$ is the  energy spacing of the SAD at
$B=0$.     
At very high magnetic fields the low-energy spectrum corresponds to
$|n=0,m\rangle$ states of the lowest Landau level (LLL).    
In this work energies are measured in effective Rydberg,    
${\cal R}=m_e^* e^4 / 2 \varepsilon^2 \hbar^2$, and lengths in   
effective Bohr radius,   $a_B = \varepsilon \hbar^2 / m_e^* e^2$.   
With $m_e^*=0.054$ $m_0$ and the dielectric constant
$\varepsilon=12.4$ we have ${\cal R}= 4.78$ meV and $a_B=
12.2$ nm.    
  
We populate the single-particle states with $N$ electrons and one   
hole.   
In the language of the creation (annihilation) operators for an   
electron $c_{i\sigma}^+$ ($c_{i\sigma}$),  
and a hole $h_{j\sigma}^+$ ($h_{j\sigma}$), the Hamiltonian 
of interacting carriers is
 \begin{eqnarray}   
  \hat{H} &=&  
  \sum_{i,\sigma} \varepsilon^e_{i} c_{i,\sigma}^+ c_{i,\sigma}    
  + {1\over2}\sum\limits_{ijkl,\sigma,\sigma'}  
       \langle ij|V_{ee}|kl\rangle c_{i,\sigma}^+ c _{j,\sigma'}^+   
                                   c_{k,\sigma'} c_{l,\sigma }    
  + \sum_{i,\sigma} \varepsilon^h_{i,\sigma} h_{i,\sigma}^+ 
  h_{i,\sigma} 
  \label{mphamil} 
  \\  
  &-&\sum\limits_{ijkl, \sigma,\sigma'}   
  \langle ij|V_{eh}|kl\rangle c_{i,\sigma}^+ h_{j,\sigma'}^+   
                              h_{k,\sigma'} c_{l,\sigma}   
  +\sum_{ij}\sum_{\sigma_1,\sigma_2,\sigma_3,\sigma_4}   
  \langle \sigma_1,\sigma_2|H^{ij}_X|\sigma_3,\sigma_4\rangle   
     c^+_{i,\sigma_1}h^+_{j,\sigma_2}h_{j,\sigma_3} c_{i,\sigma_4},  
     \nonumber  
\end{eqnarray}  
where the composite indices $i$, $j$, $k$, $l$ enumerate the 
single-particle states, i.e., $|i\rangle=|nm\rangle$, and    
$\langle ij|V_{ee}|kl\rangle$ and $\langle ij|V_{eh}|kl\rangle$ are 
respectively the electron-electron (e-e) and electron-hole (e-h) 
Coulomb matrix elements, proportional to
$V_0=\sqrt{\pi}/l^e$, with $l^e = 1/\sqrt{\Omega_h^e}$ being the
electronic oscillator length \cite{ehmatrixelements}.     
The term $V_{eh}$ describes only the direct e-h attraction,  
while  the last term in Eq.~(\ref{mphamil}) describes the   
e-h exchange. 
For orbitals $i,j$, and in the basis of electron and hole spin   
states $\{|\downarrow\Uparrow\rangle, |\uparrow\Downarrow\rangle,   
|\uparrow\Uparrow\rangle,|\downarrow\Downarrow\rangle\}$,   
$\hat{H}_{X}$  is approximated by \cite{manfred-fine}  
\begin{equation}   
  H^{ij}_X = {1\over2}\left(   
    \begin{array}{cccc}   
      \Delta_0^{ij} & \Delta_1^{ij} & 0 & 0 \\   
      \Delta_1^{ij} & \Delta_0^{ij} & 0 & 0 \\   
      0 & 0 & -\Delta_0^{ij} & \Delta_2^{ij} \\   
      0 & 0 & \Delta_2^{ij} & -\Delta_0^{ij} \\   
    \end{array}   
  \right).   
  \label{ehxhamil}
\end{equation}   
The exchange matrix elements $\Delta_{\alpha}^{ij}$ are calculated as  
$\Delta_{\alpha}^{ij}\approx \Delta_{\alpha}   
\int{d\vec{r} |\phi_{i}^e(\vec{r})|^2 |\phi_{j}^h(\vec{r})|^2 }$,   
with $\phi_{i}^e$, $\phi_{j}^h$  being the electron and hole orbitals,
and $\Delta_{\alpha}$  - material-dependent parameters.
This allows to express the elements $\Delta_{\alpha}^{ij}$ in terms of
the fundamental elements $\Delta_{\alpha}^{00}$, with $\Delta_0^{00}=
400$ $\mu$eV, $\Delta_1^{00}= 180$ $\mu$eV,  and $\Delta_2^{00}= 90$
$\mu$eV \cite{manfred-fine}.

The eigenenergies and eigenstates of the charged exciton are
calculated in the configuration-interaction approach in the basis of  
all electron-hole configurations classified by total angular momentum
and total spin $S_z$.
The valence hole couples many electronic configurations and
correlations are important, but many conclusions can be
drawn from a simplified analysis given below.  
 
Next, we use the Fermi's Golden Rule
$I(\omega) = \sum_f |\langle N-1,f|P^-|N,i\rangle|^2   
  \delta(E_{i}-E_{f}-\omega)$
to calculate the emission spectrum $I(\omega)$ from the initial 
state $|N,i\rangle$ of the charged exciton with energy $E_i$ 
to all final states $|N-1,f\rangle$ of $N-1$ electrons with energy
$E_f$, remaining in the system.
Here $P^- = \sum_j ( c_{j\uparrow}h_{j\Downarrow}+   
c_{j\downarrow}h_{j\Uparrow}) $ is the interband polarization   
operator \cite{pawelp}.    
  
We start our qualitative analysis by  mapping out the phase diagram of  
the system  of $N$ electrons and one hole as a function of the 
magnetic field. 
We start with the $\nu=2$ integer quantum Hall state   
$|\nu=2\rangle_{2M} = \left(\prod\limits_{m=0}^{M-1}   
c_{0m\uparrow}^+c_{0m\downarrow}^+\right)h^+_{00\Uparrow}|0\rangle$   
as a reference configuration of the charged exciton with $N=2M$ 
electrons.
This configuration is a ground state of the system over a finite range
of magnetic fields.
At high magnetic fields it is unstable against the edge spin flips.
The first spin-flip configuration 
$|1SF\rangle=c^+_{0M\downarrow}c_{0,M-1\uparrow}|\nu=2\rangle_{2M}$    
is an edge spin-flip state, in which one electron spin-up was removed   
from the highest occupied orbital, and placed, with spin down, on the   
next, unoccupied LLL orbital. 
At low magnetic fields the $|\nu=2\rangle$ configuration is unstable
against a center spin flip.
The spin-triplet configuration   
$|CF\rangle = c^+_{10\downarrow}c_{0,M-1\uparrow}|\nu=2\rangle_{2M}$   
stabilizes as a result of a spin-flip transfer of the electron from
the edge of the droplet to the center orbital $|10\rangle$, belonging
to the second LL.   
In a similar fashion we seek the ground-state configurations and 
energies of the charged exciton as a function of the
magnetic field.
In Fig.~\ref{fig1} we plot the lowest energies for $N=1$ (a neutral
exciton $X$) to $N=8$ (an exciton with seven charges $X7-$) with
respect to the reference energy $E_{\nu=2}(N)$ (traces for different
values of $N$ have been shifted apart for clarity).   
The horizontal plateau marks the region of magnetic fields where the
$|\nu=2\rangle_N$ configuration is the ground state of the system.   
As we move from this plateau towards the higher magnetic fields, we see 
a series of cusps marking the  spin flips.
Counting the number of spin flips allows us to read out the charge   
of the SAD with accuracy of $\pm1$ electron.  
For example, we find two spin flips for both $N=4$ and 
$N=5$.  
The two electron numbers can be further distinguished by analyzing the 
details of the emission spectrum, as discussed below.   
 

In Fig.~\ref{fig1} the first spin flip appears for all
charged excitons, while the second spin flip is first seen
for the triply charged exciton $X3-$ ($N=4$).   
The $X3-$ complex, whose energy is marked with the   
thick line, is also the first one to exhibit the center spin flip at   
low fields, and this is why  we focus on it in the rest of   
this work.   
 
Figure~\ref{fig2}(a) shows the magnetic-field evolution of the
ground-state energy of $X3-$, and the configurations dominant in the
corresponding eigenstates (diagrams at the top of the  Figure). 
The first one is the $|CF\rangle$ , the second is the $|\nu=2\rangle$ 
reference state, the third is  the $|1SF\rangle$ and
the fourth is the second spin flip $|2SF\rangle$ configuration. 

Let us now account for the electron-hole exchange Hamiltonian
$\hat{H}_X$ in a perturbative manner.
This Hamiltonian couples the members of the total electronic spin
manifold and the spin of the valence hole.
For example, for the state $|1SF\rangle$ all possible spin
configurations of the valence hole and the unpaired electrons can be
classified into two families:
$\{ |\downarrow\downarrow\Uparrow\rangle,   
|\downarrow\uparrow\Downarrow\rangle,   
|\uparrow\uparrow\Uparrow\rangle\}$   
and   
$\{|\uparrow\uparrow\Downarrow\rangle,   
|\downarrow\uparrow\Uparrow\rangle,   
|\downarrow\downarrow\Downarrow\rangle\}$.   
Both families are described by the same e-h exchange Hamiltonian matrix, 
constructed by writing the operator $\hat{H}_X$ (\ref{ehxhamil}) in
the above basis.
We diagonalize this matrix numerically to obtain three exchange-split
levels originating from the three members of the electronic spin $S=1$
manifold.
Their energies are added to the energy $E_{1SF}$.   
Upon the inclusion of the Zeeman effect, each level further splits   
into two, reflecting the differences in spin alignments of    
the two families.   
A similar analysis can be carried out for the CF and second spin-flip 
configuration $|2SF\rangle$. 
Since in the $|2SF\rangle$ state the electrons have total spin $S=2$,
we obtain families of five exchange-split levels, derived from five
members of the electronic spin $S=2$ manifold: $S_z=0,\pm1,\pm2$. 

Let us now turn to the analysis of the emission spectrum of the $X3-$
complex.
The dominant final-state configurations corresponding to each phase
are depicted in the lower part of Fig.~\ref{fig2}(a), while the resulting
calculated emission spectra are shown in Fig.~\ref{fig2}(b).  

In the recombination process, the low-magnetic field configuration
$|CF\rangle$ couples to two final states of the three electrons,
$S={3\over2}$ (shown as a spin-polarized state) at high energy and
$S={1\over2}$ at low energy. 
Each line is, in general, further split by the Zeeman energy,
reflecting the two possible spin alignments of the pair of recombining
carriers, and by the gaps arising from the e-h exchange interaction in 
the initial state. 
In Fig.~\ref{fig2}(b) we only account for the latter, setting
$g_e=g_h=0$.

The $|\nu=2\rangle$ configuration is an electronic spin singlet, not
affected by the e-h exchange.
This phase can recombine to two final states, composed of
configurations shown in the bottom part of
Fig.~\ref{fig2}(a), second diagram from the left.    
Thus, we again obtain two emission lines.  
For the initial $|\nu=2\rangle$ configuration containing the hole   
spin-down the emission spectrum is the same, except for the splitting
of the lines introduced by the Zeeman terms.   
 
In the $|1SF\rangle$ phase, the optically active final 
configuration shown in the bottom part of Fig.~\ref{fig2}(a) 
(third diagram  from the left) is mixed with three additional
configurations. 
Two of them are created by changing the position of the spin-up
electron, while the fourth one is an Auger-type excitation.   
These configurations can be grouped into one $S={3\over2}$ state
(high-energy emission line) and a group of three
$S={1\over2}$ states visible in Fig.~\ref{fig2}(b)  at lower energy.  
Due to the triplet character of the initial state, each of these lines
shows a fine structure of three maxima.
   
Finally, the only possible final-state configuration left after the   
recombination from the $|2SF\rangle$ state is the spin-polarized
configuration shown in the bottom part of Fig.~\ref{fig2}(a) in the
first diagram from the right.   
This is an $S={3\over2}$ configuration.  
Here, the inclusion of the other members of the $|2SF\rangle$   
multiplet results in several additional final states, whose energies   
are larger than $E_{1}^{2SF}$.   
In addition, each line will become split due to Zeeman and
fine-structure effects.   
Since the $S=2$ multiplet is composed of five
configurations, the fine structure of the emission line is
composed of five closely-spaced maxima.

At the magnetic fields corresponding to spin flips in
$X3-$, the energies of the adjacent phases are equal, but
the energies of the corresponding final states differ.
As a result, the emission spectrum, shown in Fig.~\ref{fig2}(b),
exhibits characteristic discontinuities at the critical magnetic
fields.
The number of discontinuities past the filling factor 2 droplet 
equals half of the number of confined electrons. 
Also, the total electronic spin of the system can be identified
in each phase from the multiplicity of the corresponding emission
line, because the different total spin couples differently to the
hole via the e-h exchange.  
   
Let us now identify the signatures of the spin flips in the emission   
spectrum of a single, mesa-selected InAs/GaAs SAD in magnetic   
fields up to 23 T.   
The structure of the sample and the details of the experiment
are described elsewhere \cite{adam-pl}.   
The excess confined charges originated from the $n^+$ doping of the
substrate, and their population was tuned by the energy of the
photoexciting laser.    
Figure~\ref{fig3}(a) shows the measured photoluminescence (PL)
spectrum which we identify as that of the 
$X3-$ complex.
In this spectrum, a single, bright maximum, detected at very low   
magnetic fields at about $1.315$ meV, exhibits a discontinuity in the
vicinity of $1$ T.   
As the field increases, we follow the single PL line up to $B\sim 8$   
T, where it splits into two maxima.   
It is not clear whether this splitting is due to the Zeeman effect   
(not seen in any other part of the spectrum), or to emission from an
excited state.
Absence of the Zeeman splitting has been reported in PL   
measurements on similar samples \cite{hifield}.   
In the region of $B=12$ T to $14$ T the spectrum broadens and appears   
to undergo a discontinuous shift to higher energies.   
At very high magnetic fields the trace consists of   
three closely spaced lines.   

We identify the low-field discontinuity of the trace as the signature   
of the spin-flip transition between the $|CF\rangle$ and the   
$|\nu=2\rangle$ phases of the $X3-$, and the discontinuity at $B\approx   
14$ T - as the onset of the first spin-flip $|1SF\rangle$ configuration.   
To confirm this interpretation, we have calculated the emission spectra of   
the $X3-$ for model parameters $\Omega_{0,e}=20$ meV, $\Omega_{0,h}=4$   
meV, $g_e=-3$ and $g_h=1$.   
The electronic $g$ factor was chosen to reproduce the critical   
magnetic field corresponding to the first spin flip,
while the chosen value of the hole $g$ factor suppresses   
the Zeeman splitting of the electron-hole pair undergoing the   
recombination.    
The calculated low-field transition energies are shown in
Fig.~\ref{fig3}(b).    
The maximum originating from the $|CF\rangle$ configuration is split   
into three peaks by the e-h exchange.   
However, the resulting energy gaps are too small to be resolved in the   
experiment, where only a single, broad maximum is seen.   
The discontinuity at $B\approx1.3$ T marks the onset of the $\nu=2$   
phase.   
The transition energies calculated in the high-field range are shown
in Fig.~\ref{fig3}(c).   
Starting with the emission from the $\nu=2$ ground state, at   
$B\approx 16$ T we find an upward discontinuity corresponding to the   
first spin flip.   
At this point, the trace acquires a three-line character - an optical   
signature of an electronic triplet, split by the e-h exchange.   
This three-line spectrum agrees well with the experimental   
trace.
 
To emphasize the fine structure seen in the emission spectra,
insets to  Fig.~\ref{fig3}(a) show the calculated and measured
emission traces for several magnetic fields. 
At $B=4$T the ground state of the $X3-$ complex is the 
spin-singlet $|\nu=2\rangle$ configuration, unaffected by the e-h 
exchange, resulting in the appearance of a single emission line. 
At $B=21.5$T the $|1SF\rangle$ configuration is the ground state. 
The e-h exchange splits the three-fold degenerate manifold of this 
triplet into three maxima, as seen in the calculations and 
experimental data. 
At even higher magnetic fields a second spin flip 
is expected, resulting in five maxima characteristic
for the $S=2$ manifold. 
This magnetic field, however, is not reached in this
experiment. 
 
In conclusion, we have demonstrated that one can optically determine
the number of carriers in a SAD by analyzing the emission
spectra of the system at high magnetic fields. 
The number of carriers determines the number of spin flips, which can
be identified as  discontinuities in the emission line.  
The total electron spin can be detected from the multiplicity of  
the emission lines. 
This opens the possibility of tuning electronic properties of
SADs with the number of carriers, with applications  
in nano-electronics, nano-spintronics, nano-optics and quantum  
information processing.


\begin{figure}[h]   
  \caption{ Energies of the charged excitons composed of $N=1$ to $8$   
    electrons and one hole as a function of the cyclotron energy.   
    In all cases the reference energy is that of the respective   
    $\nu=2$ configuration.   
    The shade of the area under each trace corresponds to the total   
    spin of the system, with darker areas denoting higher spins.   
  }   
  \label{fig1}   
\end{figure}

\begin{figure}[h]   
  \caption{ (a) Energy of the triply charged exciton as a function of   
    the magnetic field.  
    Diagrams show the initial and final state configurations   
    in each regime, while the vertical dashed lines mark the spin flips.   
    (b) The calculated emission spectrum as a function of the
    cyclotron energy for $g_e=g_h=0$ and
    with the electron-hole exchange treated perturbatively. }     
  \label{fig2}   
\end{figure}   
   
\begin{figure}[h]   
  \caption{ (a) (Color online) 
     The measured PL spectrum of the triply charged exciton   
     in a single InAs/GaAs self-assembled quantum dot.
     White lines are guides to the eye.
     Bottom panels show the emission spectra at low (b) and high (c)
     magnetic fields calculated with model parameters
     $\Omega_{0,e}=20$ meV, $\Omega_{0,h}=4$ meV, $g_e=-3$, and
     $g_h=1$.   
     Insets show the calculated and measured fine structure in the
     emission spectrum of $X3-$ for $B=4$T and $B=21.5$T (bottom
     axes correspond to the measured result, top axes - to
     the calculated maxima).  } 
  \label{fig3}   
\end{figure}

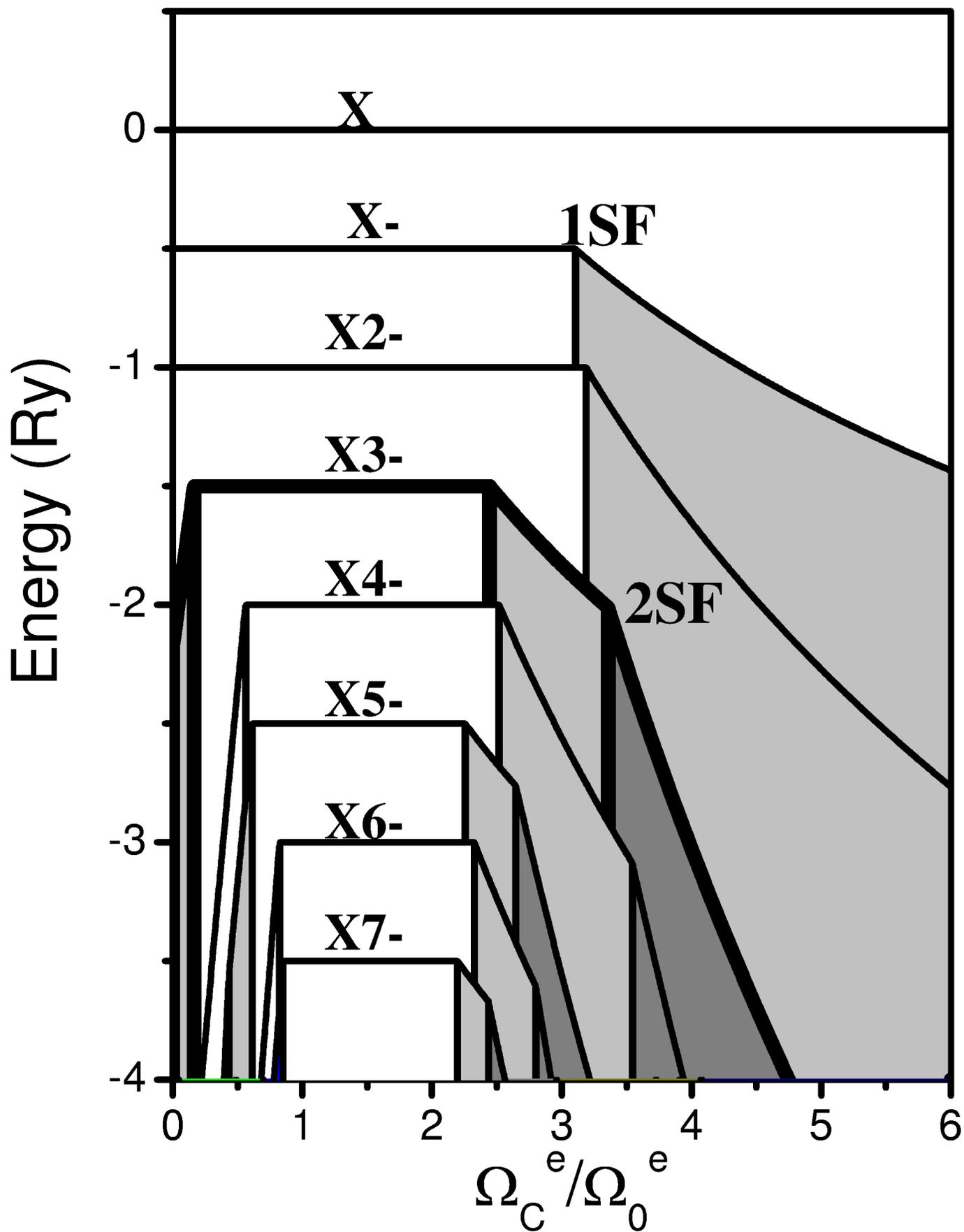
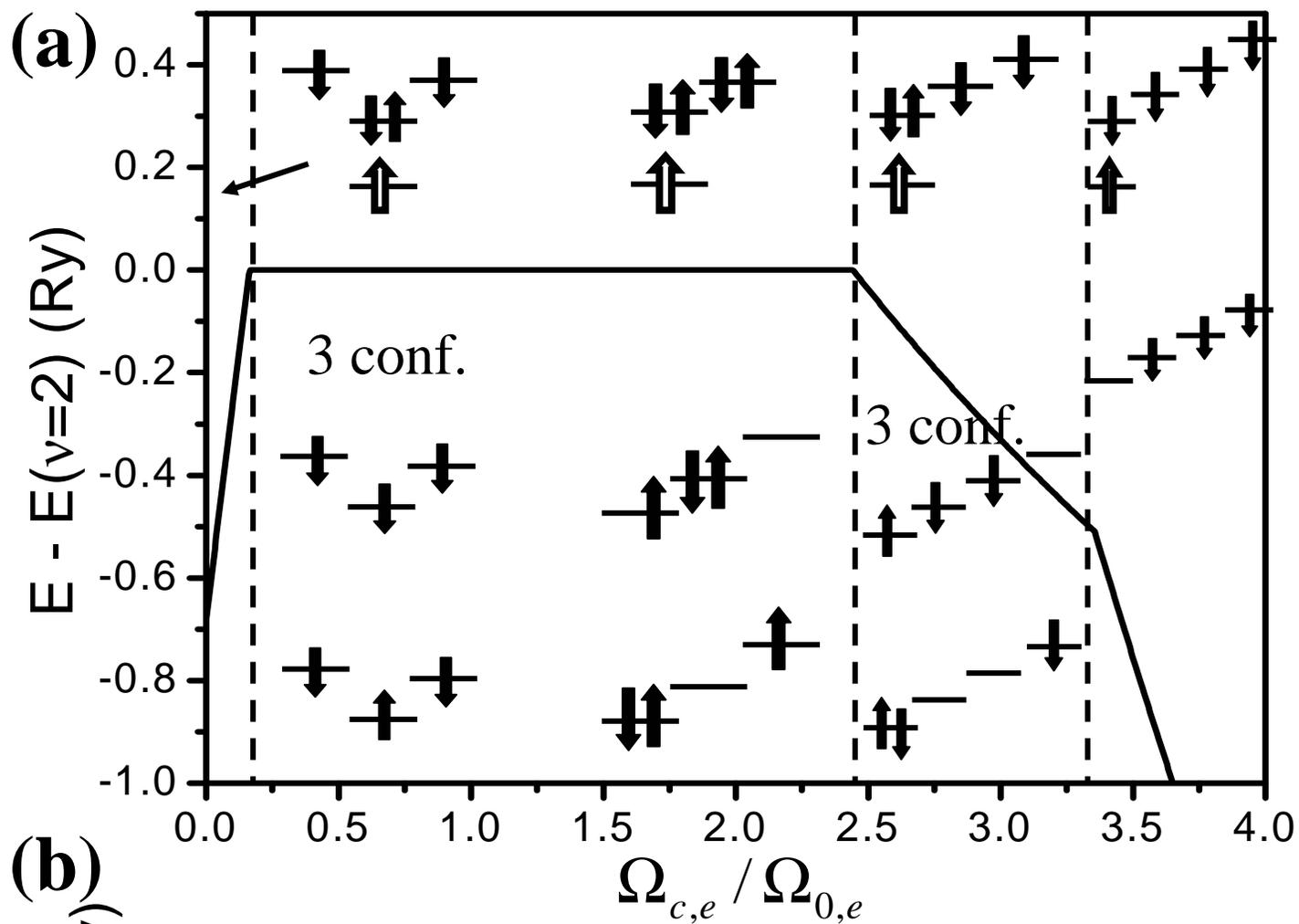
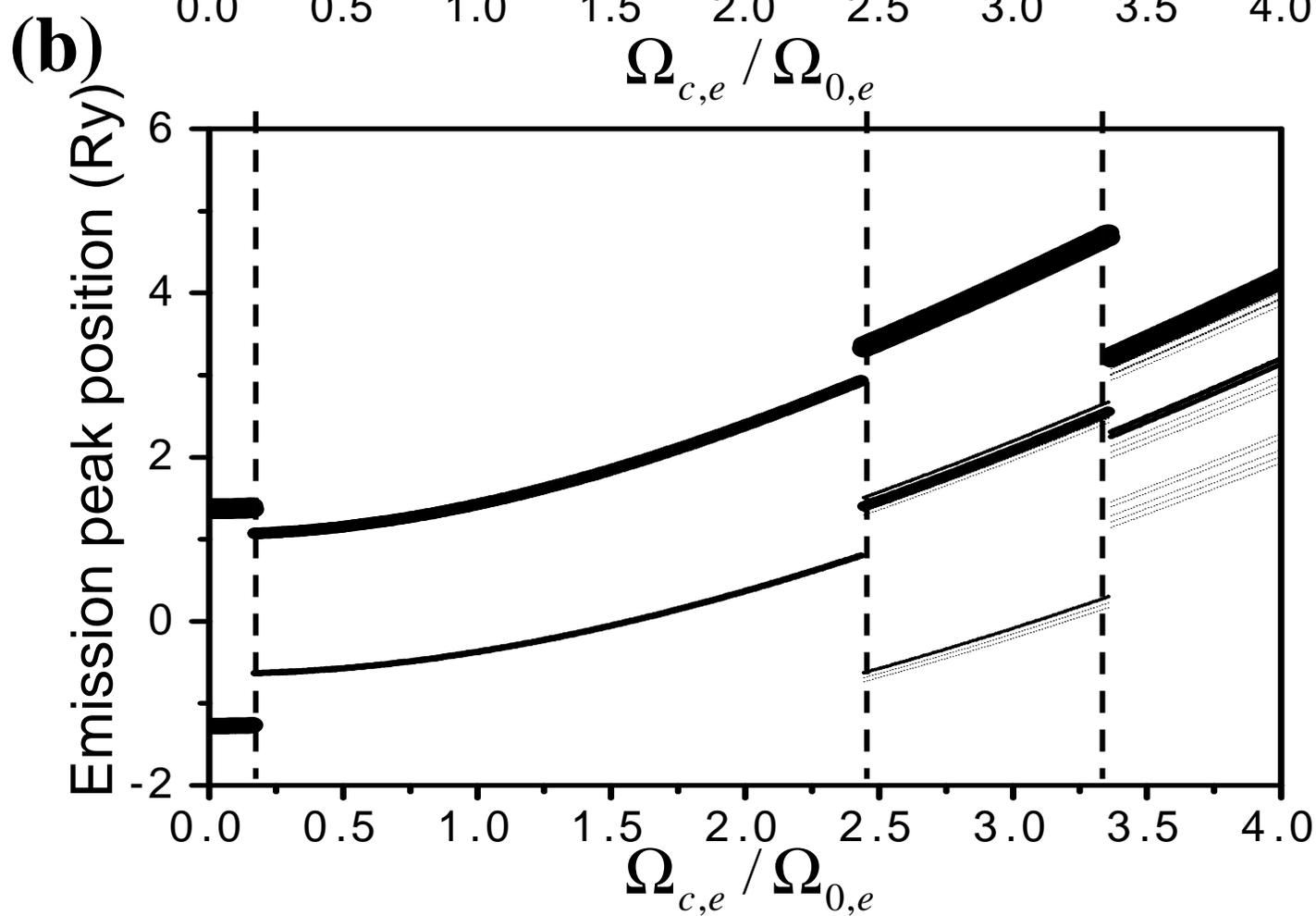
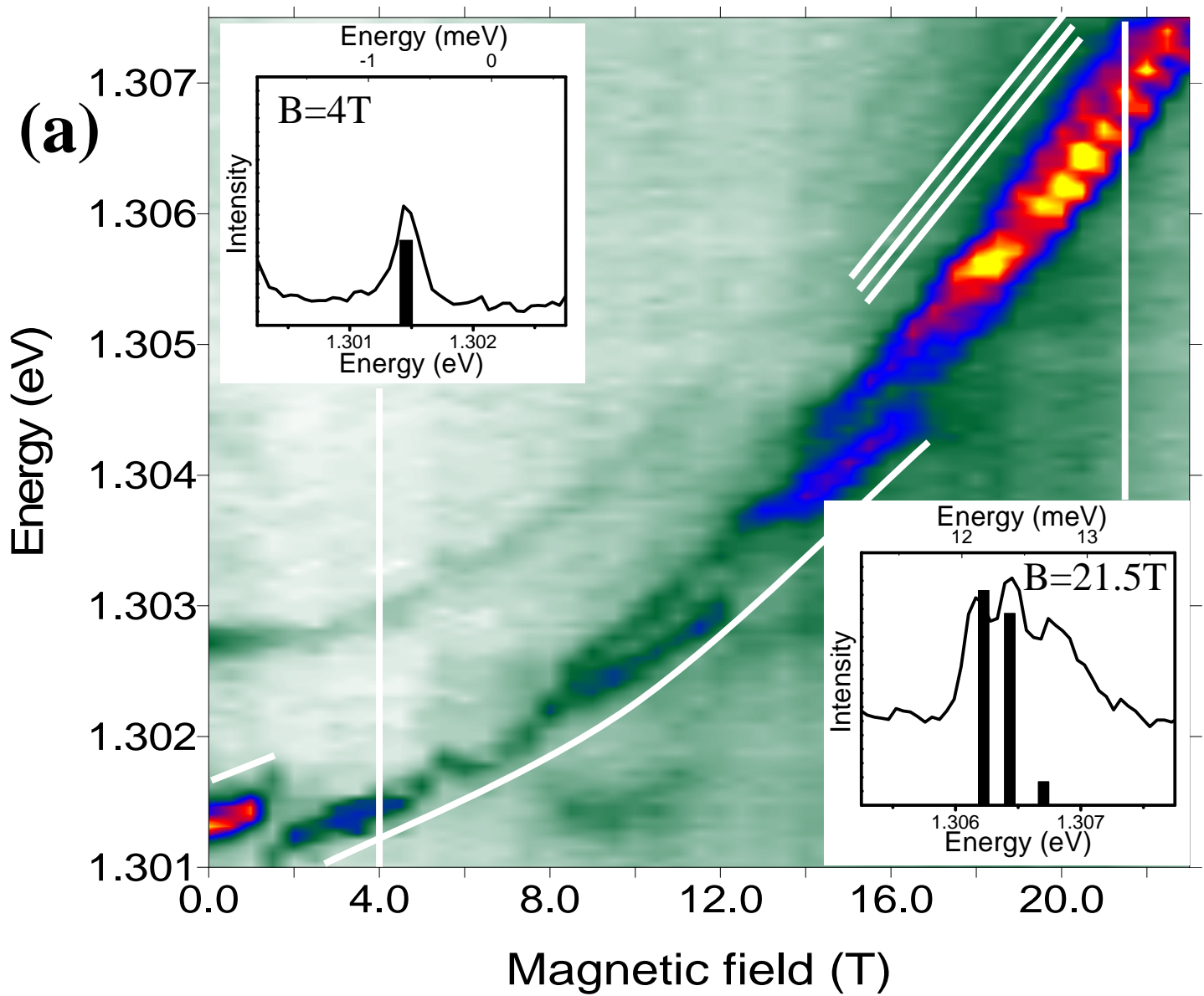
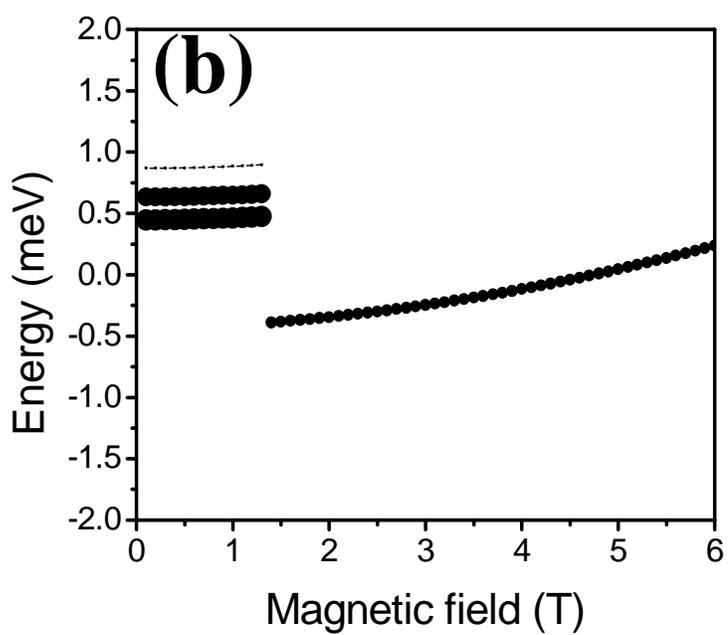
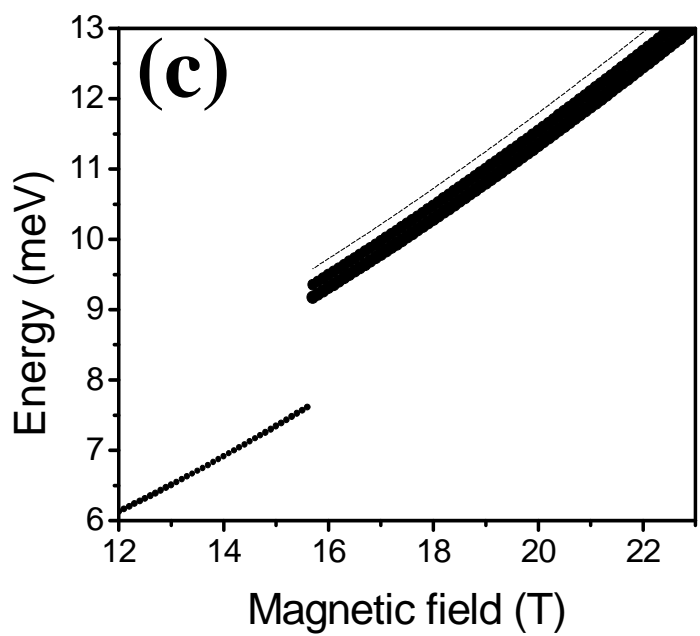
\end{document}